\documentclass[traditabstract]{aa}
\usepackage{graphicx,psfig}

\newcommand{\Tstar} {T$_{\rm{eff}}$}

\newcommand{\simless}{\mathbin{\lower 3pt\hbox
      {$\rlap{\raise 5pt\hbox{$\char'074$}}\mathchar"7218$}}} 
\newcommand{\simgreat}{\mathbin{\lower 3pt\hbox
     {$\rlap{\raise 5pt\hbox{$\char'076$}}\mathchar"7218$}}} 

\begin{document}

\newcommand{\hi}{\ion{H}{i}~}
\newcommand{\hii}{\ion{H}{ii}~}

   \title{Dust grain growth in $\rho$-Ophiuchi protoplanetary disks}

   \author{L. Ricci         \inst{1}
\and          L. Testi          \inst{1} 
\and A. Natta \inst{2} \and K. J. Brooks \inst{3}}


   \institute{  European Southern Observatory,
   Karl-Schwarzschild-Strasse 2, D-85748 Garching, Germany
            \and
                     INAF - Osservatorio Astrofisico di Arcetri, Largo Fermi 5, I-50125 Firenze, Italy
           \and Australia Telescope National Facility, P.O. Box 76, Epping, NSW 1710, Australia}

   \date{Received May 25 2010/ Accepted Aug 5 2010}

   \titlerunning{Dust grain growth in the $\rho$-Oph disks}

   \authorrunning{Ricci et al.}

\abstract{ We present new ATCA observations at 3.3 mm of 27 young stellar objects in the $\rho-$Oph young cluster. 25 of these sources have been detected. We analyze the sub-millimeter and millimeter SED for a subsample of 17 isolated class II protoplanetary disks and derive constraints on the grain growth and total dust mass in the disk outer regions. All the disks in our sample show a mm slope of the SED which is significantly shallower than the one observed for the ISM at these long wavelengths. This indicates that 1) class II disks in Ophiuchus host grains grown to mm/cm-sizes in their outer regions, 2) formation of mm/cm-sized pebbles is a fast process and 3) a mechanism halting or slowing down the inward radial drift of solid particles is required to explain the data. These findings are consistent with previous results in other star forming regions. We compare the dust properties of this sample with those of a uniformly selected sample in Taurus-Auriga and find no statistical evidence of any difference in terms of grain growth between the two regions. Finally, in our sample the mm slope of the SED is not found to correlate with indicators of grain growth to micron sizes in the surface layers of the inner disk.}



\maketitle


\section{Introduction}

Circumstellar disks around pre-main sequence (PMS) stars are the common birth places of planetary systems. In order to build up large bodies such as giant planets a huge growth of solid particles by more than 12 orders of magnitude in size has to occur starting from submicron-sized microscopic dust grains as those typically found in the interstellar medium (ISM; Mathis et al.~\cite{Mat77}). The first stages of this process of grain growth are characterized by the dynamical interaction between gas and dust, leading to collisions between the solid particles and finally coagulation (see Beckwith et al.~\cite{Bec00}, Dominik et al.~\cite{Dom07}, Natta et al.~\cite{Nat07}).

Evidence for the presence of micron-sized dust grains in protoplanetary disks has been provided by the inspection of ISO and Spitzer spectra for the silicate feature at about 10~$\mu$m (e.g. Bouwman et al.~\cite{Bou01}, Van Boeckel et al.~\cite{Van03}, Kessler-Silacci et al.~\cite{Kes06}). However infrared observations can only probe the uppermost surface layers of the disk, while to investigate the properties of dust in the disk midplane, where planet formation is expected to take place, observations at longer wavelengths are needed. If the long-wave emission is optically thin the spectral index of the SED at these wavelengths can be related to the spectral index of the dust opacity $\beta$ ($\kappa_{\rm{dust}} \propto \lambda^{-\beta}$), and $\beta$-values lower than $\sim$ 1 are naturally interpreted in terms of grain growth (Draine~\cite{Dra06}). Beckwith \& Sargent~(\cite{Bec91}) were the first to show that T Tauri Stars (TTS) have SEDs at submillimeter wavelengths which are typically shallower than what found for the ISM, suggesting that dust grains as large as at least 1 mm are present in the disk midplane. However these single-dish observations did not have a good enough angular resolution to spatially resolve the disks, and the same data could in principle be explained also by optically thick emission from compact disks with unprocessed, ISM-like grains.

Wilner et al.~(\cite{Wil00}) and Testi et al.~(\cite{Tes03}) resolved the disks around TW Hya and CQ Tau PMS stars at 7~mm with the Very Large Array (VLA), confirming that the long-wave emission from these disks reveals the presence of cm-sized pebbles in the disk midplane. More recently Rodmann et al.~(\cite{Rod06}) and Lommen et al.~(\cite{Lom07}) found evidence of grain growth to $\sim$ mm/cm-sized pebbles for about ten T Tauri stars in the Taurus-Auriga, Chamaeleon and Lupus star forming regions (SFRs). Lommen et al.~(\cite{Lom10}) have combined new and literature data for disks from five different SFRs (Taurus-Auriga, Lupus, Chamaeleon, Corona Australis and Serpens) and reported a tentative correlation between the mm slope of the SED and the strength of the 10-$\mu$m silicate feature, possibly suggesting that inner and outer disk evolve simultaneously in terms of dust grain growth.

So far, most of the observations carried out at long wavelengths have targeted the brightest sources.
Ricci et al.~(\cite{Ric10}, hereafter R10) have started to probe fainter disks with new sensitive data at 3~mm obtained with the Plateau de Bure Interferometer in Taurus and analyzed a sample of 21 isolated class II Young Stellar Objects (YSOs) in this SFR. For all the disks in this sample they found neither evidence for time evolution of dust grain growth nor significant relations between dust properties and stellar ones were found.

In this paper we perform the same analysis presented in R10, but on a sample of disks in another SFR, i.e. $\rho-$Ophiuchi. In Sect.~2 we present new 3~mm data for 27 $\rho-$Oph YSOs obtained with the Australia Telescope Compact Array (ATCA\footnote{The Australia Telescope Compact Array is part of the Australia Telescope which is funded by the Commonwealth of Australia for operation as a National Facility managed by CSIRO.}) and its new Compact Array Broadband Backend (CABB). In Sect.~3 we describe the properties of the sub-sample used for the analysis. The method adopted for deriving the dust properties, namely grain growth and dust mass, is the same used in R10 and it is described in Sect.~4. The results of the analysis are described in Sect.~5, whereas Sect.~6 summarizes the main findings of our work.

\section{New ATCA observations}
\label{sec:obs}

We observed 27 YSOs in the $\rho-$Oph star forming region at 3.2944~mm with ATCA and the new CABB digital filter bank. Targets were chosen by being class II (most of them) or ``flat spectrum'' YSOs, with a detected flux at 1.3mm higher than 20 mJy in order to have good chances for detection at $\sim$ 3.3~mm with a few hours at most of integration time on-source. 
 
The observations were carried out between the end of April and the beginning of July 2009, with the ATCA array in the H168 and H75 configurations respectively.
The obtained FWHM of the synthesized beam of $\sim 3-7''$  did not allow us to spatially resolve any of our sources, and so a 2D-gaussian model was used to estimate the flux density in the continuum at 3.3~mm. To obtain the best sensitivity in the continuum we set the correlator to cover the full 2-GHz CABB effective bandwidth on all the five antennas available at $\sim$ 3~mm.
Each source has been observed one or more times, with a total integration time on-source ranging between about 40 minutes and 3 hours. 

The complex gain calibration was done on the calibrators 1622-297 and 1730-130, and science targets were typically observed for 10 minutes and then spaced out with 2 minutes on the gain calibrator.
The absolute calibration was done with Uranus except for a few times in which 1921-293 was used and then its flux was always cross-calibrated with Uranus the day after. In general, the baselines that resolved out Uranus were not used for flux calibration\footnote{For the data reduction we used the MIRIAD package.}. In this work, we assume a 1$\sigma-$uncertainty on the calibrated flux of about 25$\%$.    

The results of the observations are reported in Table~\ref{tab:obs}. In particular we have detected 25 out of 27 targeted YSOs. The detected sources have fluxes ranging from about 1.5 and 48.8~mJy at 3.3~mm.
For the two undetected sources, IRS 33 and GY 284, we obtained an upper limit for the flux at 3.3~mm of 1.2~mJy. These sources have been previously detected by Motte et al.~(\cite{Mot98}) using the IRAM 30m single-dish telescope at 1.3~mm. Although their reported fluxes are relatively large ($105 \pm 20$~mJy and $130 \pm 10$~mJy for IRS 33 and GY 284 respectively) the emission for both the sources appears to be spatially resolved by a beam with a FWHM of about 11 arcsec. This indicates that a significant fraction of the collected flux at 1.3~mm comes from an extended envelope surrounding the circumstellar disk. In order to get a reliable estimate for the flux at $\sim$ 1~mm from the circumstellar disk only, interferometric observations at $\sim$ 1~mm are needed to filter out the contaminating emission from the envelope. Since these are not available in the literature we do not include IRS 33 and GY 284 in our sample discussed in the next section.

\section{Sample}
\label{sec:sample}

In this section we describe some properties of the sample considered for our analysis\footnote{Note that the final sample used in the analysis does not comprise all the targets of the new ATCA observations described in Sect. \ref{sec:obs}.}. In Sect. \ref{sec:selection_criteria} we outline the selection criteria adopted to select our sample, Sect. \ref{sec:completeness} describes the sample completeness level, and finally in Sect. \ref{sec:stars} we derive the stellar parameters.

\subsection{Selection criteria}
\label{sec:selection_criteria}

We used the same selection criteria adopted in R10:

\begin{enumerate}
 \item class II YSOs as catalogued in Andrews \& Williams~(\cite{And07a}) from the infrared SED to avoid contamination of (sub-)mm fluxes by a residual envelope; 
 \item central stars need to be well characterized through optical-NIR spectroscopic/photometric data to obtain self-consistent disk SED models;
 \item YSOs with at least one detection in the 0.45 $\leq \lambda \leq$ 1.3 mm spectral range\footnote{Note that in R10 a slightly narrower spectral range of 0.45 $\leq \lambda \leq$ 0.85 mm was chosen. We adopted a broader spectral range to include in our sample the sources YLW 16c and WSB 52, for which no observations at $\lambda <$ 1.3 mm have been carried out so far. The different uncertainties for the spectral slopes of these sources, due to a shorter spectral leverage, are properly taken into account in the analysis.} other than being observed through the new ATCA observations described in Sect.~2, to have a good sampling of the (sub-)mm SED;
 \item no evidence of stellar companions with projected physical separation between 5 and 500 AU to avoid tidal interactions that may alter the outer disk structure; adopting for all the YSOs in our sample the current estimate of $\sim$ 130 pc for the distance to the $\rho-$Oph star forming region (Wilking et al.~\cite{Wil08}, Lombardi et al.~\cite{Lom08}), the range in projected phyical separation translates into a range in angular separation of $\sim 0.05 - 4''$. 
\end{enumerate}  

Among the 25 YSOs detected with ATCA at 3.3~mm, 17 satisfy these selection criteria and they constitute the sample of our analysis. They are listed in Table~\ref{tab:sample}.   

\subsection{Completeness}
\label{sec:completeness}

\begin{figure*}[htbp!]
 \centering
\begin{tabular}{cc}
\includegraphics[scale=0.45]{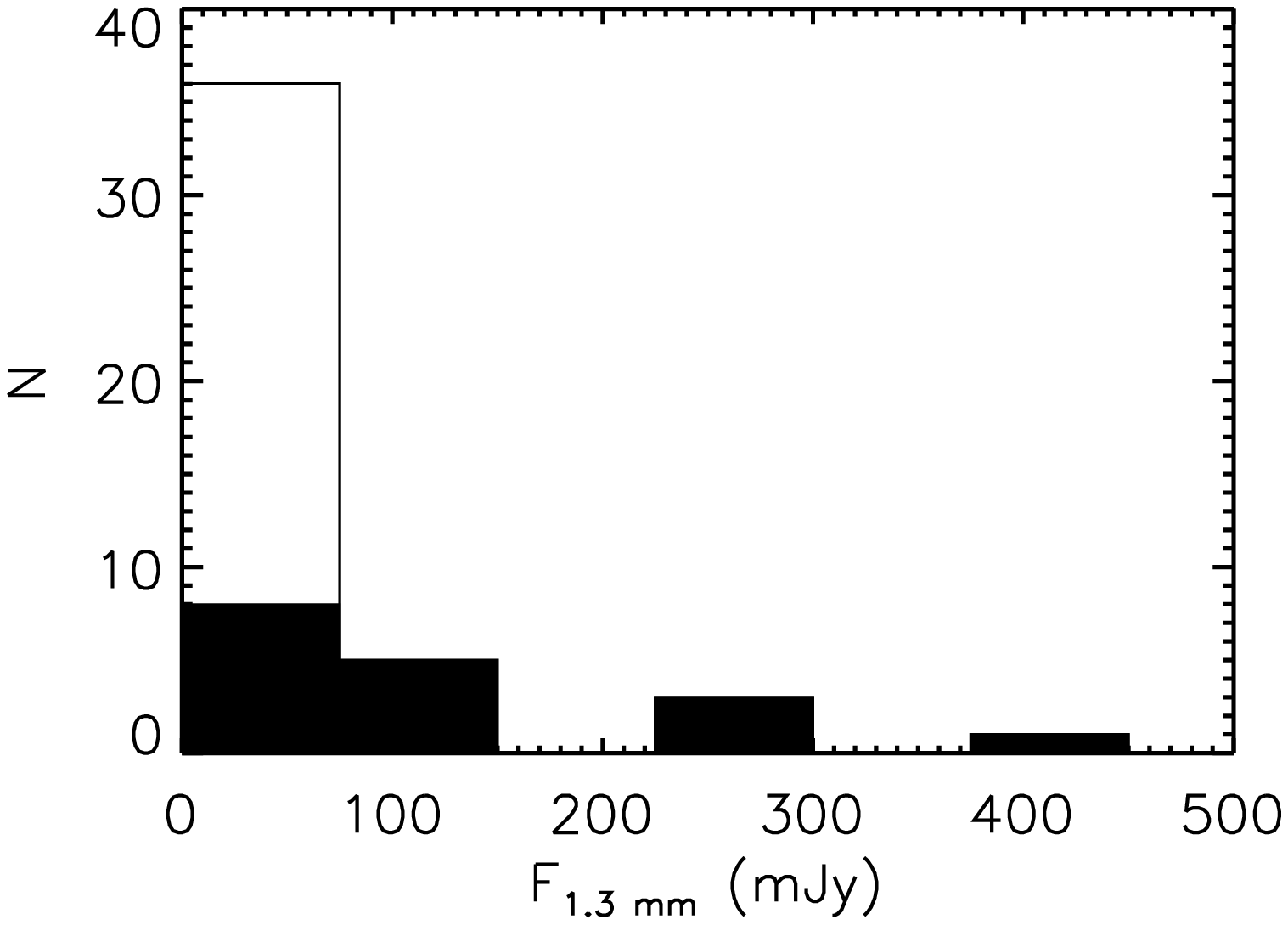} &
\includegraphics[scale=0.45]{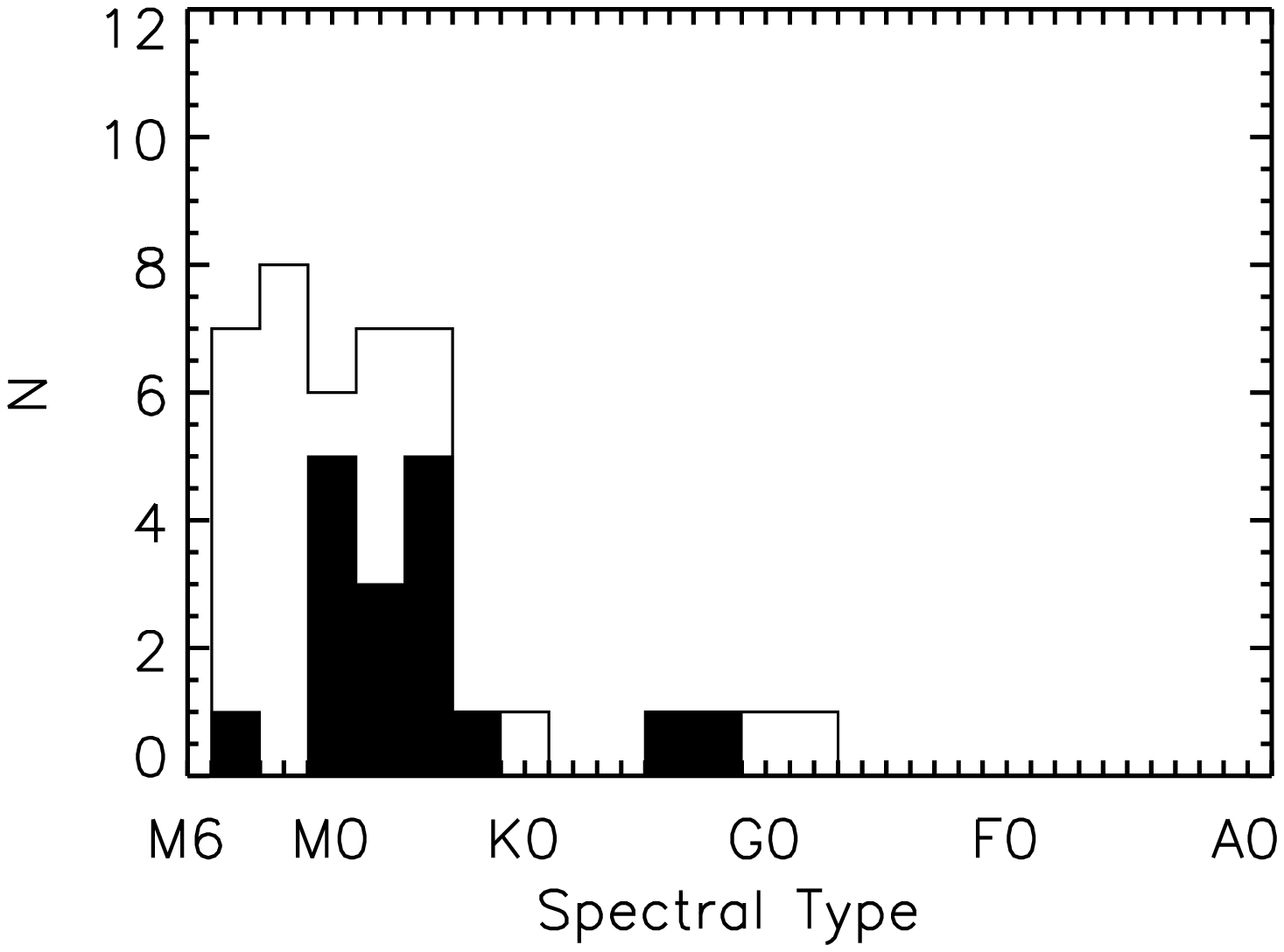} \\

 \end{tabular}
 \caption{Histograms higlighting some properties of our selected sample. In both the histograms our sample is represented by black columns, while the total columns (black$+$white) include all the class II YSOs from the Andrews \& Williams~(\cite{And07a}) catalogue with no evidence of stellar companions in the $0.05-4.0''$ interval in angular separation, from which our sample has been selected (see Sect. \ref{sec:selection_criteria}). Left) Distribution of the fluxes at 1.3 mm, including upper limits; Right) Distribution of stellar spectral type (see Sect. \ref{sec:stars}).}

 \label{fig:completeness}
\end{figure*}

The two histograms in Fig.~\ref{fig:completeness} show the distribution of the ``isolated'', i.e. satisfying the selection criterion (4) in Sect.~\ref{sec:selection_criteria}, class II YSOs from the Andrews \& Williams~(\cite{And07a}) catalogue. The histogram on the left side shows that our sample comprises all the isolated class II YSOs with $F_{\rm{1.3mm}} > 75$ mJy, while for the fainter objects the completeness level reduces to 22\% ($8/36$).

In terms of the stellar properties, our sample includes 53\% ($16/30$) of the isolated PMS stars with stellar types equal to or earlier than M1, correspondent to a stellar mass larger than $\sim$ 0.5 $M_{\odot}$ for a $\sim 1$ Myr old PMS star according to the Palla \& Stahler~(\cite{Pal99}) models\footnote{Note that this completeness level is very close to the one obtained in R10 in the Taurus-Auriga star forming region, i.e. $\sim 58\%$ of the isolated class II YSOs with stellar masses larger than $\sim 0.4 \ M_{\odot}$.} (see Fig.~\ref{fig:hrd}). Our sample is much less complete for PMS stars with later spectral types (or smaller stellar mass): only 7\% of the known PMS stars with a mass smaller than $\sim 0.4 \ M_{\odot}$ is in our sample ($1/15$), and this has to be considered as an upper limit since the Andrews \& Williams~(\cite{And07a}) catalogue does not comprise class II YSOs with spectral types later than M5. This is due to the sensitivity limits of the current (sub-)mm facilities to detect emission from circumstellar material around very-low mass PMS stars.

\subsection{Stellar properties}
\label{sec:stars}

To constrain the stellar properties we used the same method as described in R10. We estimated the stellar effective temperatures by converting the adopted spectral types (from Andrews \& Williams~\cite{And07a} and references therein\footnote{For only two cases in our sample, i.e. EL 20 and IRS 41, the spectral types are not available in Andrews \& Williams~(\cite{And07a}). For these PMS stars we adopted the stellar types from Natta et al.~(\cite{Nat06}) which are based on near infrared broad band photometry.}) with the dwarf temperature scale of Schmidt-Kaler~(\cite{Sch82}) for spectral types earlier than M0 and the intermediate temperature scale of Luhman~(\cite{Luh99}) for spectral types equal to or later than M0.

Stellar luminosities were computed from the 2MASS $J$-band flux (Cutri et al.~\cite{Cut03}) after 
calculating the extinction of each object by dereddening the $J-H$ and $H-K_{s}$ colors to the locus observed for Classical T Tauri stars (Meyer et al.~\cite{Mey97}), and adopting the Cardelli et al.~(\cite{Car89}) extinction law with $R_{V} = 4.2$, which is appropriate for $\rho-$Oph\footnote{For IRS 41, for which the 2MASS $J$-band flux is not available, we adopted the luminosities as derived by Natta et al.~(\cite{Nat06}), after correcting them by a multiplicative factor $(130$ pc$/150$ pc$)^2$ to account for the different adopted distance of 150 pc in the Natta et al. paper.}.

\begin{figure}
 \centering
 \resizebox{\hsize}{!}{\includegraphics{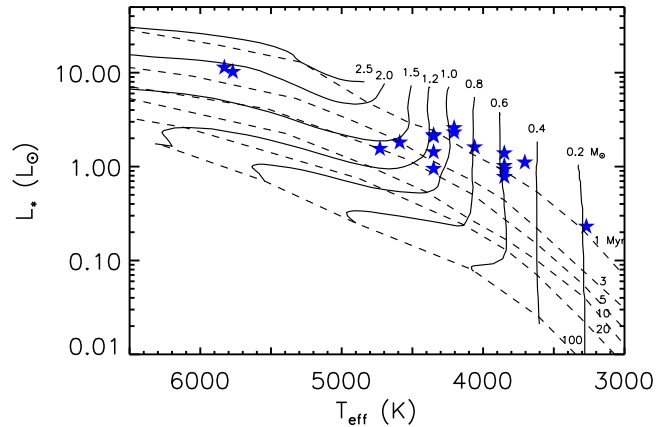}}
 \caption{H-R diagram for the sources of our sample. The dashed and solid lines represent the isochrones and evolutionary tracks respectively from the Palla \& Stahler~(\cite{Pal99}) PMS evolutionary models. In this diagram the evolutionary tracks start from an age of $0.1$ Myr. The uncertainties, not shown in the plot, are typically $\sim 0.08-0.13$ in $\log L_{\star}$ and $\sim 100-150$ K in temperature.}
 \label{fig:hrd}
\end{figure}

Luminosities and effective temperatures were converted into stellar masses and ages by using the Palla \& Stahler~(\cite{Pal99}) models of PMS stars, as done in R10 (see Fig.~\ref{fig:hrd}). According to these models the ranges spanned by our sample go from about 0.2 $M_{\odot}$ (WSB 60) to 1.9 $M_{\odot}$ (SR 21) in mass and from about 0.7 Myr (GSS 26, EL 24, YLW 16c) to 6.6 Myr (SR 21) in age.

The stellar parameters are reported in Table~\ref{tab:sample}.

\begin{table*}

\centering \caption{ Summary of the 3.3mm ATCA observations. } \vskip 0.1cm
\begin{tabular}{lccrrccc}
\hline \hline
\\
Object  & $\alpha$ & $\delta$ & $F_{\nu}$ & rms   & Comments$^{\rm{b}}$ & Ref. & Other names \vspace{1mm} \\
        &  (J2000) & (J2000)  & (mJy)     & (mJy) &   &    &  \\
\\
\hline
\\
AS 205A  & 16:11:31.4    & -18:38:26.0 & 27.2  & 1.2 & triple (1.5'', sb) & 1 & HBC 254 \vspace{1mm} \\
SR 4     & 16:25:56.1    & -24:20:48.3 & 4.4   & 0.4 &   & & AS 206  \vspace{1mm}  \\
GSS 26   & 16:26:10.3    & -24:20:54.9 & 24.2  & 0.9 &   & &     \vspace{1mm}  \\
EL 20    & 16:26:18.9    & -24:28:20.2 & 7.3   & 0.5 &   & & VSSG 1  \vspace{1mm}  \\
LFAM 1   & 16:26:21.7    & -24:22:50.8 & 17.5  & 0.6 & flat spectrum  & 2 &          \vspace{1mm}  \\
DoAr 24E & 16:26:23.4    & -24:21:00.7 & 8.3   & 0.4 & binary (2.0'') & 1 &  GSS 31  \vspace{1mm}  \\
DoAr 25  & 16:26:23.7    & -24:43:14.1 & 25.0  & 0.6 &   & & WSB 29  \vspace{1mm}  \\
EL 24    & 16:26:24.1    & -24:16:14.0 & 48.8  & 0.7 &   & & WSB 31  \vspace{1mm}  \\
EL 27    & 16:26:45.0    & -24:23:08.2 & 38.7  & 0.6 &   & & GSS 39  \vspace{1mm}  \\
WL 18    & 16:26:49.0    & -24:38:25.7 & 3.1   & 0.5 & binary (3.6'') & 3 &  GY 129  \vspace{1mm}  \\
SR 24S   & 16:26:58.5    & -24:45:37.1 & 26.6  & 0.8 & flat spectrum, triple (5.2'', 0.2'') & 2,4,5  &  HBC 262 \vspace{1mm}  \\
SR 21    & 16:27:10.2    & -24:19:12.9 & 4.2   & 0.4 &  & &  EL 30   \vspace{1mm}  \\
IRS 33$^{\rm{a}}$ & 16:27:14.5  & -24:26:46.1 & $< 1.2$ & 0.4 & & & GY 236  \vspace{1mm}  \\
IRS 41   & 16:27:19.3    & -24:28:44.4 & 6.2   & 1.2 &  & &  WL 3    \vspace{1mm}  \\
CRBR 85  & 16:27:24.7    & -24:41:03.2 & 1.5   & 0.3 & envelope  & 6 &        \vspace{1mm}  \\
YLW 16c  & 16:27:26.5    & -24:39:23.4 & 6.5   & 0.4 &   & & GY 262  \vspace{1mm}  \\
GY 284$^{\rm{a}}$ & 16:27:30.8  & -24:24:56.0 & $< 1.2$ & 0.4 & & &          \vspace{1mm}  \\ 
IRS 49   & 16:27:38.3    & -24:36:58.8 & 4.4   & 0.4 &   & & GY 308  \vspace{1mm}  \\
DoAr 33  & 16:27:39.0    & -23:58:19.1 & 3.7   & 0.3 &   & & WSB 53  \vspace{1mm}  \\
WSB 52   & 16:27:39.5    & -24:39:15.9 & 10.2  & 0.5 &   & & GY 314  \vspace{1mm}  \\
IRS 51   & 16:27:39.8    & -24:43:15.0 & 12.7  & 0.8 & binary (1.6'') & 3 &  GY 315  \vspace{1mm}  \\
WSB 60   & 16:28:16.5    & -24:36:58.0 & 15.3  & 0.5 &   & & YLW 58  \vspace{1mm}  \\
SR 13    & 16:28:45.3    & -24:28:19.2 & 10.0  & 0.6 & binary (0.4'') & 3 &  HBC 266 \vspace{1mm}  \\
DoAr 44  & 16:31:33.5    & -24:27:37.7 & 10.4  & 0.5 &   & & HBC 268 \vspace{1mm}  \\
RNO 90   & 16:34:09.2    & -15:48:16.9 & 7.6   & 0.5 &   & & HBC 649 \vspace{1mm}  \\
Wa Oph 6 & 16:48:45.6    & -14:16:36.0 & 10.3  & 0.4 &   & & HBC 653 \vspace{1mm}  \\
AS 209   & 16:49:15.3    & -14:22:08.6 & 17.5  & 0.5 &   & & HBC 270 \vspace{2mm}  \\

\hline
\end{tabular}
\begin{flushleft}
$a$) For this undetected source the ($\alpha$,$\delta$) coordinates are from 2MASS (Cutri et al.~\cite{Cut03});

$b$) Reason why the source has not been considered in the analysis (see Sect. \ref{sec:selection_criteria}); enclosed in parenthesis there are the angular separation(s) between the stellar companions (``sb''$=$ spectroscopic companion); 

$c$)References: (1) McCabe et al.~(\cite{McC06}); Andrews \& Williams~(\cite{And07a}); (3) Ratzka et al.~(\cite{Rat05}); (4) Reipurth \& Zinnecker~(\cite{Rei93}); (5) Simon et al.~(\cite{Sim95}); (6) McClure et al.~(\cite{McC10}).
\end{flushleft}

\label{tab:obs}

\end{table*}

\begin{table*}

\centering \caption{ Stellar properties of the considered sample. } \vskip 0.1cm
\begin{tabular}{lcrrrrr}
\hline \hline
\\
Object  & ST  & \Tstar & $L_{\star}$ & $R_{\star}$ & $M_{\star}$ & Age \vspace{1mm} \\
&     &   (K)  &  ($L_{\odot}$)    &  ($R_{\odot}$) & ($M_{\odot}$) & (Myr)   \\
\\
\hline
\\

SR 4     &   K5     &   4350  &  2.17 & 2.58 & 1.14 & 1.07 \vspace{1mm} \\
GSS 26   &   M0     &   3850  &  1.39 & 2.64 & 0.56 & 0.54 \vspace{1mm} \\
EL 20    &   M0     &   3850  &  0.93 & 2.16 & 0.62 & 1.07 \vspace{1mm} \\
DoAr 25  &   K5     &   4350  &  1.43 & 2.10 & 1.12 & 2.09 \vspace{1mm} \\
EL 24    &   K6     &   4205  &  2.58 & 3.01 & 0.96 & 0.58 \vspace{1mm} \\
EL 27    &   M0     &   3850  &  0.78 & 1.98 & 0.58 & 1.23 \vspace{1mm} \\
SR 21    &   G3     &   5830  & 11.38 & 3.29 & 1.97 & 2.21 \vspace{1mm} \\
IRS 41   &   K7     &   4060  &  1.61 & 2.55 & 0.80 & 0.76 \vspace{1mm} \\
YLW 16c  &   M1     &   3705  &  1.11 & 2.55 & 0.48 & 0.54 \vspace{1mm} \\
IRS 49   &   M0     &   3850  &  1.02 & 2.26 & 0.64 & 1.00 \vspace{1mm} \\
DoAr 33  &   K4     &   4590  &  1.81 & 2.12 & 1.44 & 2.88 \vspace{1mm} \\
WSB 52   &   K5     &   4350  &  0.95 & 1.71 & 1.04 & 4.17 \vspace{1mm} \\
WSB 60   &   M4     &   3270  &  0.23 & 1.49 & 0.20 & 0.93 \vspace{1mm} \\
DoAr 44  &   K3     &   4730  &  1.55 & 1.85 & 1.29 & 5.13 \vspace{1mm}  \\
RNO 90   &   G5     &   5770  & 10.24 & 3.19 & 1.87 & 2.32 \vspace{1mm} \\
Wa Oph 6 &   K6     &   4205  &  2.32 & 2.86 & 0.98 & 0.71 \vspace{1mm} \\
AS 209   &   K5     &   4350  &  2.11 & 2.55 & 1.18 & 1.23 \vspace{1mm} \\

\hline
\end{tabular}

\label{tab:sample}

\end{table*}

\section{Analysis}
\label{sec:analysis}

In order to constrain the dust properties in the disk outer regions we fitted the sub-mm/mm SED with    
two-layer (i.e. disk surface and midplane) models of flared disks heated by the radiation of the central star (Chiang \& Goldreich~\cite{Chi97}, Dullemond et al.~\cite{Dul01}).
In Sect.~\ref{sec:disk_structure} and \ref{sec:dust_opacity} we describe the parameters which are needed to define a two-layer disk model. 
In Sect.~4.3 we then discuss which are the physical quantities of these models that can be constrained by our analysis.

\subsection{Disk structure}
\label{sec:disk_structure}

In order to characterize a model of the disk estimates for some stellar physical quantities (bolometric luminosity $L_{\star}$, effective temperature $T_{\rm{eff}}$ and mass $M_{\star}$), plus some information on the disk structure and on the the dust opacity are needed. As for the stellar parameters we used the values listed in Table~\ref{tab:sample}.
Regarding the disk structure the only relevant parameters for our analysis are the disk outer radius $R_{\rm{out}}$ and the parameters $\Sigma_{\rm{dust,1}}$ and $p$ which define a power-law surface density for the dust component: 

\begin{equation}
\Sigma_{\rm{dust}}(R) = \Sigma_{\rm{dust,1}}\left (\frac{R}{\rm{1\ AU}}\right )^{-p}, 
\end{equation}
radially truncated at $R_{\rm{out}}$. For the disk outer radii we adopted the intervals listed in Column (3) of Table~\ref{tab:fits_results}. For the seven disks in our sample which have been mapped by high angular resolution observations at sub-mm/mm wavelengths presented in Andrews \& Williams~(\cite{And07b}) we considered intervals roughly centered on their best-fit values. Three sources (EL 20, WSB 52, DoAr 44) have been observed and spatially resolvedby recent observations carried out by Andrews et al.~(\cite{And09}) with the SMA at 0.87~mm: these disks show evidence of an outer radius larger than 100~AU, but an accurate estimate of the outer radius $R_{\rm{out}}$ could not be derived, since in order to fit the interferometric visibilities they adopted the self-similar solution for a viscous disk (Lynden-Bell \& Pringle~\cite{Lyn74}) rather than a truncated power-law disk. For the seven sources in our sample which have not been mapped yet we adopted a fiducial interval for $R_{\rm{out}}$ of $100-300$ AU, which comprises nearly all the disks outer radii as found for the 24 Taurus and Ophiucus circumstellar disks in Andrews \& Williams~(\cite{And07b}). Based on the results of this same work we adopt for all the sources in our sample an interval for the surface density power-law index $p$ which goes from 0.5 to 1.5. 

An important thing to keep in mind here is that with these values for the outer radius of the disk the dust emission at the long wavelengths considered in this paper turns out to be dominated by the optically thin outer disk regions. This has two important consequences for our discussion. The first one is that constraints on dust properties such as dust grain sizes and dust mass can be derived from the continuum emission at sub-mm/mm wavelengths. If a disk is much more compact than the ones which have been commonly mapped so far, i.e. if $R_{\rm{out}} \simless 20-30$ AU, then its emission would be dominated by the denser inner regions which are optically thick even at these long wavelengths, and no information on the dust properties could be inferred by the observed continuum. The second consequence is that the disk inclination is not a relevant parameter for our analysis, except only for the case of a nearly edge-on disk that however would make the central PMS star invisible in the optical.

\subsection{Dust opacity}
\label{sec:dust_opacity}

To calculate the dust opacity we adopted the same dust grain model taken in R10, i.e. porous composite spherical grains made of astronomical silicates (optical constants from Weingartner \& Draine~\cite{Wei01}), carbonaceous materials (Zubko et al.~\cite{Zub96}) and water ices (Warren~\cite{War84}) with fractional abundances from a simplification of the model used in Pollack et al.~(\cite{Pol94}) and a volume fraction for vacuum of $\approx 30\%$.
In the disk surface and midplane we consider a dust grain population with a grain size number density 
throughout all the disk

\begin{equation}
n(a) \propto a^{-q}   
\end{equation}
between $a_{\rm{min}}$ and $a_{\rm{max}}$ ($a$ here is the grain radius). In this paper we adopt two different grain size distributions in the disk surface and midplane, since both observations and theory predict larger grains in the midplane. In particular, although we consider in both the surface and midplane a small value of $a_{\rm{min}} \approx 0.1\ \mu$m, for the maximum grain size $a_{\rm{max}}$ in the disk surface we consider a value only slightly larger than $a_{\rm{min}}$\footnote{We also consider the same value of $q$ for both the surface and the midplane. However, contrary to the dust in the midplane where $q$ is a relevant parameter, in the disk surface the adopted $q-$value is practically non influential since $a_{\rm{max}}$ is only slightly larger than $a_{\rm{min}}$}, whereas $a_{\rm{max}}$ has been left free to vary in the disk midplane. As done in R10 we consider in this work four possible values for the $q-$parameter: 2.5, 3.0, 3.5, 4.0.

\subsection{Method}

After having set the disk outer radius $R_{\rm{out}}$
and the power-law index $p$ of the surface density profile, the two-layer models can be used to fit the sub-mm/mm SED of circumstellar disks to constrain dust properties in the disk midplane\footnote{Note that this is true only in the case in which the dust emission at long wavelengths comes mostly from the optically thin disk outer regions, as described at the end of Sect.~\ref{sec:disk_structure}}., in particular the spectral index of the dust opacity $\beta$ between two mm-wavelengths (1 and 3~mm in the case of this paper) and the product $M_{\rm{dust}} \times \kappa_{\rm{1mm}}$ between the mass in dust and the dust opacity at 1~mm. Except for very low values of $\beta$, which cannot be explained by large values of the power-law index $q$ of the grain size number-density (see discussion in Sect.~5.1), every $\beta$-value can be reproduced by different ($q$, $a_{\rm{max}}$) couples. Furthermore, at a fixed $q$, the precise value of the maximum grain size $a_{\rm{max}}$ correspondent to a certain derived $\beta$ depends strongly on the model that one adopts for the dust. The only robust conclusion which is valid for all the reasonable models of dust analyzed so far is that $\beta$-values lower than the value found for the ISM ($\beta_{\rm{ISM}} \sim 1.7$) can be obtained only with dust populations in which grains as large as at least $\sim$ 1 mm are present (see e.g. Natta et al.~\cite{Nat07}). For these reasons in the rest of the paper we will rarely refer to the maximum grain size $a_{\rm{max}}$, whereas we will more frequently use $\beta$ as our proxy for grain growth.

In order to derive an estimate for $\kappa_{\rm{1mm}}$ and thus constrain the dust mass $M_{\rm{dust}}$, one has to adopt a certain model for the dust grain which provide a family of functions $\kappa_{\rm{1mm}}(\beta)$ labeled with the $q$-parameter. The $\kappa_{\rm{1mm}}(\beta)$ functions considered in this paper have been obtained using the dust model presented in Sect.~\ref{sec:dust_opacity} and they are the same shown in Figure~3 in R10. 

\begin{figure*}[htbp!]
 \centering
 \includegraphics[scale=0.8]{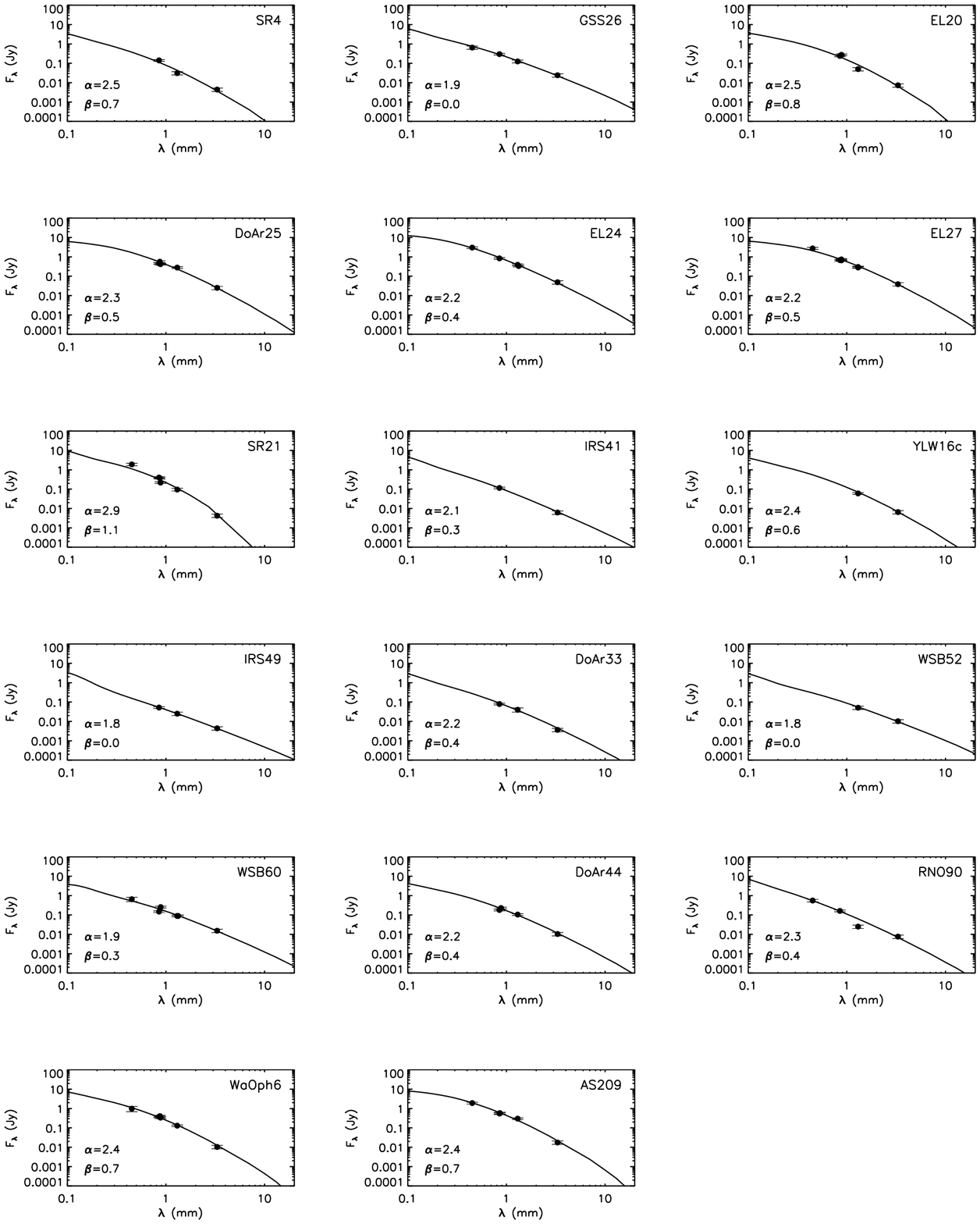}
 \caption{Sub-mm/mm SED fits for the 17 sources in our sample with the two-layer flared disk models (solid lines). The errorbars in the plots take into account an uncertainty of 10\% on the absolute flux scale at wavelengths shorter than 3 mm, and of 25\% at 3 mm. The fitting values of the spectral index $\alpha$ and of the dust opacity spectral index $\beta$ between 1 and 3 mm are indicated in the bottom left corner of each plot. The value of the adopted surface density power-law index $p$ is 1, the adopted value for the outer disk radius is the central value of the $R_{\rm{out}}-$interval listed in Table~\ref{tab:fits_results}, whereas the value for the power-law index of the grain size number-density $q$ is 3 except for four disks (GSS 26, YLW 16c, IRS 49, WSB 52) for which models with $q=3$ cannot explain the data (see Table~\ref{tab:fits_results}). For these sources models with a $q-$value of 2.5 are shown.}
 \label{fig:SED_fits}
\end{figure*}

Figure~\ref{fig:SED_fits} reports the best fit flared disk model overplotted to the sub-mm/mm data for each disk in our sample\footnote{Note that for a few disks (SR 4, EL 20, RNO 90) the flux at $\sim$ 1~mm falls below the model line. Although this could be due to problems in the observations, another possible reason is that the 3~mm-flux for these sources is contaminated by free-free emission. If this was the case, the derived $\beta$-values for these disks would be only lower-limits. Observations at longer wavelengths, where free-free dominates the emission, are needed to constrain its possible contribution at 3~mm.}. Since as explained in Sect.~4.1 for each disk we have adopted, instead of a single value, an interval of possible values for $R_{\rm{out}}$ and $p$, the uncertainty on these parameters translates into an uncertainty on the quantities derived by fitting the sub-mm/mm SED, i.e. $\beta$ and $M_{\rm{dust}} \times \kappa_{\rm{1mm}}$. Adding this contribution to the uncertainties in the observational data, the total absolute uncertainties are approximately 0.4 for $\beta$ and a factor of $\approx 3-4$ for $M_{\rm{dust}} \times \kappa_{\rm{1mm}}$.

\section{Results}
\label{sec:results}

The results of the SED-fitting procedure are listed in Table~3. Here we discuss them in terms of dust grain growth (Sect.~5.1) and mass in dust (Sect.~5.2) for our sample of protoplanetary disks.

\subsection{Grain growth}
\label{sec:grain_growth}

Information on the level of dust grain growth in the outer regions of protoplanetary disks comes from the analysis of the spectral index of the dust opacity at (sub-)millimeter wavelengths $\beta$, which reflects the spectral index of the disk SED at these long wavelengths $\alpha$. In particular for a completely optically thin disk in the Rayleigh-Jeans regime $\beta = \alpha -2$, whereas if emission from the optically thick disk inner regions (i.e. $R \simless 20-30$ AU) and deviations of the emitted spectrum from the Rayleigh-Jeans regime are taken into account, as done for our analysis, $\beta \simgreat \alpha -2$. In the limit case of a completely optically thick disk even at these long wavelengths the SED spectral index $\alpha$ becomes independent on $\beta$ and so no information on grain growth would be obtainable. In Columns (5) and (6) of Table~3 the constrained values of $\alpha$ and $\beta$ between 1 and 3~mm are reported\footnote{Note that the difference $\alpha-\beta$ turns out to be in the range $1.6-1.9$. This (small) discrepancy from the value of 2 is primarly due to the low temperature of the outer disk midplane and the consequent deviation from the Rayleigh-Jeans regime of the mm-wave emission.}.

Figure~4 shows the SED spectral index $\alpha_{\rm{1-3mm}}$ plotted against the observed flux at 3.3~mm for all the sources in our sample. The spanned range in $\alpha_{\rm{1-3mm}}$ is $1.8-2.9$, and there is no clear evidence of any correlation between the two plotted quantities. The fact that all the disks in our sample show a (sub-)mm spectral index which is shallower than the one found for the ISM, i.e. $\alpha < \alpha_{\rm{ISM}}$, brings evidence of grain growth from an initial ISM-like dust population for all the disks in our sample\footnote{This sentence is strictly valid for the ten disks in our sample which have been mapped so far; for the seven disks which have not been mapped yet this sentence is valid only if the underlying assumption that their spatial extension is not so small that most of the mm emission is optically thick ($R_{\rm{out}} \simless 20-30$ AU) holds true.}.

\begin{figure}[htbp!]
 \centering
 \includegraphics[scale=0.54]{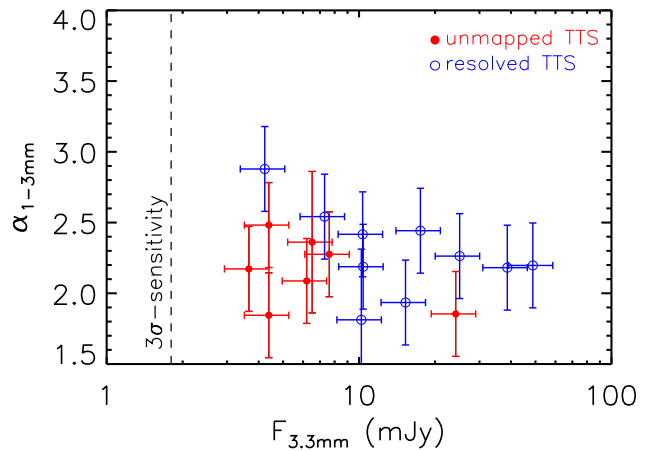}
 \caption{Spectral index $\alpha$ between 1 and 3~mm from the model best fits shown in Figure~3 plotted against the observed flux at 3.3~mm. Open blue points reprensent the spatially resolved disks, whereas the filled red points are for the unmapped ones. The dashed vertical line shows the mean 3$\sigma$-sensitivity of our ATCA observations, i.e. about 1.8~mJy.}
\end{figure}

In Figure~5 we report the histogram of the derived distribution of $\beta$-values. All the disks show $\beta < \beta_{\rm{ISM}} \sim 1.7$, indicative of dust grain growth to at least mm-sizes. Note that for 9 out of the 17 disks $\beta \simless 0.8$ (considering a $1\sigma$-uncertainty on $\beta$ of about 0.4) and for them the data are not consistent with the MRN-value of 3.5 for the power-law index $q$ of the dust grain size distribution in the ISM (Mathis et al.~\cite{Mat77}): for these disks $q-$values as low as $2.5-3$ are needed (see discussion in R10).

\begin{figure}[htbp!]
 \centering
 \includegraphics[scale=0.45]{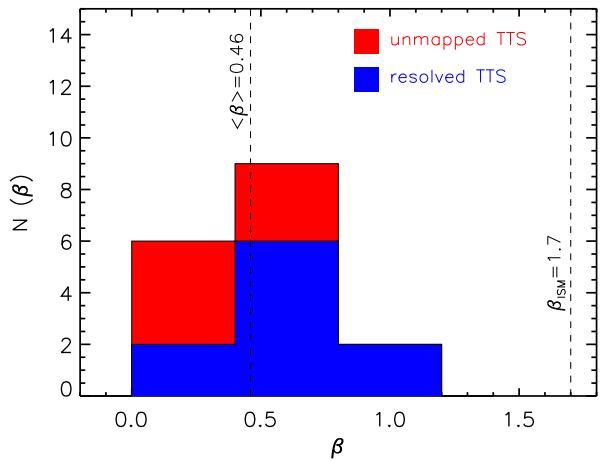}
 \caption{Distribution of the dust opacity spectral index $\beta$ for the disks in our sample. In blue the values for the spatially resolved disks are indicated, whereas the red is for the unmapped ones. The average $\beta$ value for all the sources in our sample and the value of $\sim 1.7$ for the ISM dust are indicated as dashed vertical lines.}
\end{figure}

The mean value is $<\beta> = 0.46 \pm 0.08$, which is consistent within 1$\sigma$ with the one obtained for the Taurus sample discussed in R10 ($<\beta>_{\rm{Tau}}=0.6 \pm 0.06$). In order to make a statistical comparison of the dust grain growth between the Taurus sample and the one presented in this paper in the $\rho-$Oph star forming region we performed a two-sample KS test. The probability that the two samples have $\beta-$values drawn from the same distribution is $\approx 22\%$. The hypothesis that the samples in Taurus and Ophiuchus have the same $\beta-$distribution cannot thus be rejected at the $95\%$ of confidence level.

In Figure~6 we plot the (sub-)mm spectral index of the dust opacity $\beta$ against the stellar age for YSOs in different evolutionary stages: the 38 class II disks in Ophiuchus and Taurus presented here and in R10 respectively, and a sample of 15 less evolved class 0 YSOs\footnote{Note that for these objects there are no robust age-estimates.} in Taurus, Perseus, and isolated from the PROSAC survey (Joergensen et al.~\cite{Joe07}) and from Kwon et al.~(\cite{Kwo09}). Contrary to the class II disks, for all the class 0 objects the values of $\beta$ have been obtained using the approximated $\beta = \alpha - 2$ relation (see caption of Figure~6). As described before, this relation gives only a lower-limit for $\beta$, and this probably explains why for many of these sources the derived $\beta$ is negative. A more sophisticated analysis is needed to get more robust estimates of $\beta$ by taking into account self-consistently deviations from the above relation as due to departures from the Rayleigh-Jeans regime of the emission (expecially for these cold sources) and to marginally optically thick emission typically associated to the compact structure forming the disk. However, the low values of $\beta$ obtained for nearly all these sources appear to show evidence for dust grain growth to $\sim$ mm-sizes (see Figure 3 in R10) already in the earliest stages of star formation. Ormel et al.~(\cite{Orm09}) have recently investigated the effects of dust coagulation and fragmentation onto the dust size distribution in molecular cloud cores. They found that grain sizes close to $\sim$ 1~mm can be formed if cloud lifetimes are not restricted to free-fall times but rather support mechanisms like e.g. ambipolar diffusion are present and if freeze-out of ice has commenced. According to their simulations ice-coated grains can grow to sizes of $\sim 0.3- 8$~mm in one ambipolar diffusion timescale at densities of $n = 10^{5} - 10^{7}$~cm$^{-3}$, which are typical of the inner regions of molecular cores. Dust grain growth to $\sim$ mm-sizes can thus be a process accompanying the very first phases of star formation.

\begin{figure}[htbp!]
 \centering
 \includegraphics[scale=0.52]{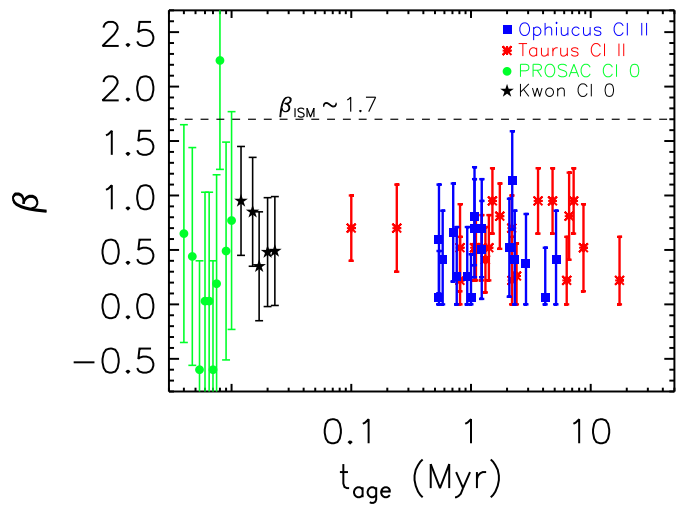}
 \caption{Dust opacity spectral index $\beta$ versus age: relationship between the dust opacity spectral index $\beta$ and the estimated stellar age obtained as in Sect.~3.3 for the class II YSOs. Note that for the class 0 YSOs (green circles and black stars) the position in the x-axis of the plot is only indicative, since no robust age estimates can be derived for these sources. Blue squares represent the sample of class II YSOs presented in this paper, red asterisks are the Taurus-Auriga class II disks from R10, green circles are the class 0 YSOs from the PROSAC survey (Joergensen et al.~\cite{Joe07}), black stars are class 0 YSOs from Kwon et al.~(\cite{Kwo09}). The $\beta$-value for the ISM dust is indicated as a dashed horizontal line. For all the class 0 YSOs the values of $\beta$ have been obtained using the $\beta=\alpha - 2$ relation, where $\alpha$ is the spectral index between the CARMA total fluxes at 1.3 and 2.7~mm for the Kwon et al. sample, and between the fluxes collected at 0.85 and 1.3~mm with the SMA array at baselines longer than 40~k$\lambda$ for the PROSAC sample. This criterion was chosen by the authors to minimize the contribution from the extended envelope to the total emission.}
\end{figure}

Figure~6 shows also that there is no relation between the dust opacity spectral index and the stellar age for the class II disks: grains as large as $\sim$ 1~mm appears to be present in the outer regions of disks throughout all the class II evolutionary stage. This is in contrast with the short timescales of inward radial drift expected for $\sim$ mm/cm-sized grains in the outer disk as a consequence of the dust interaction with the gas component. In order to explain the retention of large dust grains in these outer regions some mechanisms which may halt the drift of solid particles, e.g. local pressure maxima due to turbulent vortices or spiral density waves, have been invoked. Birnstiel et al.~(\cite{Bir10b}) have compared the observed fluxes at millimeter wavelengths for the disks samples described in R10 and in this paper with predictions of dust evolution models accounting for coagulation and fragmentation. They showed that, if radial drift of solid particles is completely suppressed, a grain size distribution at the steady-state (due to a balance between coagulation and fragmentation) can explain the mm-wave emission of the brightest disks. The observed flux of the fainter disks are instead typically overpredicted even by more than one order of magnitude. These discrepancies may be explained by considering in the disk models a dust reduction due to radial drift at a reduced rate\footnote{The radial drift, other than decreasing the amount of dust in the disk and thus decreasing the flux at millimeter wavelengths, is more efficient for mm/cm-sized pebbles than for smaller grains in the outer disk. For this reason, if one wants to explain the low values of the mm-spectral indeces only a reduced rate of radial drift (from that expected theoretically) can be invoked.} or during an earlier evolutionary time or due to efficient conversion of dust into larger, unseen bodies (see Birnstiel et al.~\cite{Bir10b} for more details). Observations of these faint disks can thus help us to determine 
which mechanisms play a major role for the dynamics and evolution of large grains in the outer regions of disks.

\begin{figure}[htbp!]
 \centering
 \includegraphics[scale=0.47]{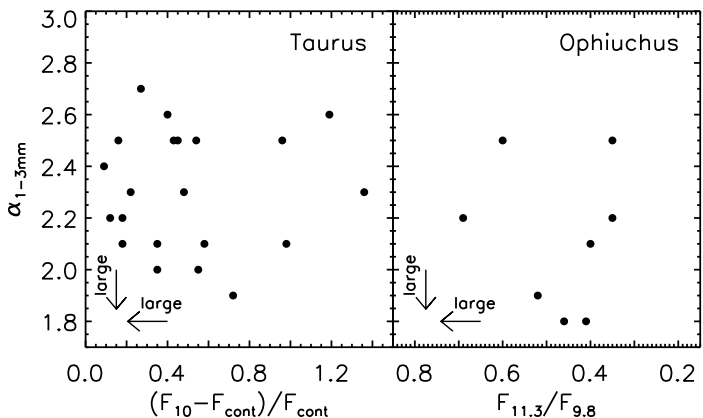}
 \caption{SED spectral index between 1 and 3~mm versus indicators of grain growth from the 10-$\mu$m silicate feature. Left) SED slope between 1 and 3~mm vs the ratio between the 10-$\mu$m line flux and the continuum as derived from Spitzer/IRS observations by Furlan et al.~(\cite{Fur06}) for the sample of Taurus-Auriga class II YSOs described in R10. Right) SED slope between 1 and 3~mm vs the ratio between the flux at 11.3 and 9.8~$\mu$m as derived from Spitzer/IRS observations by McClure et al.~(\cite{McC10}); the plotted points are all the class II YSOs described in this paper with a value of $F_{11.3}$/$F_{9.8}$ reported in McClure et al.. In the two plots the arrows in the lower left corners show toward which direction the values of the plotted quantities are indicative of the presence of largest grains.}
\end{figure}

Figure~7 shows the spectral index between 1 and 3~mm in Taurus (left) and Ophiuchus (right) plotted against two indicators of dust processing from the silicate feature observed at about 10~$\mu$m (see e.g. Kessler-Silacci et al.~\cite{Kes06}). Data for the 10~$\mu$m silicate feature for the two samples in Taurus and Ophiuchus come from the literature and refer to different indicators (see caption of Figure~7) since we could not find the same indicator for both the regions. However Lommen et al.~(\cite{Lom10}) showed that these two indicators correlate well and so they can be both used to probe the growth of grains from interstellar, submicron sizes to sizes of several microns in the disk surface layers. In particular low values of the ratio between the 10~$\mu$m line flux and the continuum, $(F_{10}-F_{\rm{cont}})/F_{\rm{cont}}$, and large values of the ratio between the fluxes at 11.3 and 9.8~$\mu$m, $F_{11.3}/F_{9.8}$, are interpreted in terms of grain growth to micron sizes (Bouwman et al.~\cite{Bou01}, Kessler-Silacci et al.~\cite{Kes06}). Very recently Lommen et al.~(\cite{Lom10}) have found a tentative correlation between these quantities for a sample of about 30 TTS and Herbig-Ae/Be systems spread over five different star forming regions, including Taurus-Auriga. Since the mm slope of the SED probes grains in the disk outer regions whereas the 10-$\mu$m silicate feature is sensitive to grains in the inner regions, the tentative correlation may indicate a parallel evolution of the inner and outer disk in terms of dust grain growth. However, in this work we do not find any correlation neither for the sample in Taurus nor for the one in Ophiuchus. Note that in the case of Ophiuchus the sample with literature data for the 10-$\mu$m silicate feature is limited to eight disks only. In Taurus our sample comprises nine of the eleven disks considered by Lommen et al., and for two of them (AA Tau and GM Aur) our derived values of $\alpha_{\rm{1-3mm}}$ are not consistent with and lower than the values used by Lommen et al.\footnote{This discrepancy is probably due to a different choice of the literature data used for the derivation of $\alpha$. In the case of R10 all the data obtained at sub-mm and mm wavelengths have been used.}. Note also that the absence of such a correlation would not be too surprising since the physical mechanisms which are responsible for the observed values of the grain growth indicators are different: the presence of mm/cm-sized pebbles in the midplane of the outer disk is mainly due to coagulation processes and mechanisms which trap these pebbles in the outer disk, whereas the presence of $\mu$m-sized grains in the surface layers of the inner disk is probably regulated by fragmentation of larger solid particles and a balance between settling and turbulence mixing which keeps these relatively small grains in the uppermost layers of the disk. Further observations with more sensitive telescopes in the future are needed to provide necessary the necessary statistics to better investigate the possible relation between dust grains in the outer and inner disk regions and to possibly constrain the processes of radial mixing and vertical settling.  

\subsection{Disk mass}
\label{sec:disk_mass}

Table~3 lists the derived dust masses for our sample of disks detected at 3~mm. As already discussed in R10, the inferred dust mass depends, at a given chemical composition and porosity for the dust grain, on the assumed value of the power-law index $q$ of the grain size distribution. This dependence, which is due to the different millimeter dust opacities obtained for different values of $q$ (see R10), is particularly strong for relatively low values of $\beta$. For example, in the case of WSB 60, with $\beta \approx 0.3$, $M_{\rm{dust}}^{q=3}$ is larger than $M_{\rm{dust}}^{q=2.5}$ by a factor of about 10.

\begin{table*}
\centering \caption{ Disk properties. } \vskip 0.1cm
\begin{tabular}{lcccccccc}
\hline \hline
\\
Object name & $R_{\rm{out}}$ (AW07) & $R_{\rm{out}}-$interval & $\alpha$ & $\beta$ & $M_{\rm{dust}}\times\kappa_{\rm{1mm}}$ & $M_{\rm{dust}}^{q=2.5}$ & $M_{\rm{dust}}^{q=3}$ & $M_{\rm{dust}}^{q=3.5}$    \vspace{1mm} \\
            &   (AU)      &  (AU)   &    &     &  ($M_{\odot} \times$ cm$^2$g$^{-1}$) & ($M_{\odot}$) & ($M_{\odot}$) & ($M_{\odot}$) \vspace{1mm} \\
(1) & (2) & (3) & (4) & (5) & (6) & (7) & (8) & (9)            \\

\\
\hline
\\

\underline{SR 4} & ... &  100$-$300   & 2.5 &  0.7  & $1.3\cdot10^{-4}$ & $2.1\cdot10^{-5}$ & $2.4\cdot10^{-5}$ & ...           \vspace{1mm} \\
\underline{GSS 26} & ... & 100$-$300  & 1.9 & 0.0  & $3.5\cdot10^{-4}$ & $2.3\cdot10^{-3}$ &  ...  &  ...           \vspace{1mm} \\
EL 20              & ... & 100$-$300  & 2.5 & 0.8  & $3.2\cdot10^{-4}$ & $4.3\cdot10^{-5}$ & $4.5\cdot10^{-5}$ & $1.9\cdot10^{-4}$          \vspace{1mm} \\
DoAr 25           & 200 &  100$-$300  & 2.3 & 0.5  & $8.0\cdot10^{-4}$ & $1.8\cdot10^{-4}$ & $2.6\cdot10^{-4}$ & ...           \vspace{1mm} \\
EL 24             &  175 & 75$-$275   & 2.2 & 0.4  & $9.9\cdot10^{-4}$ & $2.9\cdot10^{-4}$ & $6.3\cdot10^{-4}$ & ...          \vspace{1mm} \\
EL 27             &  275 & 175$-$375  & 2.2 & 0.5  & $1.5\cdot10^{-3}$ & $3.5\cdot10^{-4}$ & $5.6\cdot10^{-4}$ & ...          \vspace{1mm} \\
SR 21             &  600 & 500$-$700  & 2.9 & 1.1  & $5.3\cdot10^{-4}$ & $4.9\cdot10^{-5}$ & $4.5\cdot10^{-5}$ & $ 5.5\cdot10^{-5}$          \vspace{1mm} \\
\underline{IRS 41}  & ... & 100$-$300 & 2.1 & 0.3  & $1.3\cdot10^{-4}$ & $6.6\cdot10^{-5}$ & $6.8\cdot10^{-4}$ & ...          \vspace{1mm} \\
\underline{YLW 16c} & ... & 100$-$300 & 2.4 & 0.0  & $2.3\cdot10^{-4}$ & $4.3\cdot10^{-5}$ & $5.6\cdot10^{-5}$ & ...        \vspace{1mm} \\
\underline{IRS 49}  & ... & 100$-$300 & 1.8 & 0.0  & $3.9\cdot10^{-5}$ & $2.2\cdot10^{-3}$ & ... & ...          \vspace{1mm}  \\
\underline{DoAr 33} & ... & 100$-$300 & 2.2 & 0.4  & $1.2\cdot10^{-4}$ & $3.4\cdot10^{-5}$ & $9.9\cdot10^{-5}$ & ...          \vspace{1mm}  \\
WSB 52              & ... & 100$-$300 & 1.8 & 0.0  & $1.4\cdot10^{-5}$ & $2.6\cdot10^{-3}$ & ... & ...          \vspace{1mm} \\
WSB 60              & 350 & 250$-$450 & 1.9 & 0.3  & $5.6\cdot10^{-4}$ & $2.9\cdot10^{-4}$ & $3.0\cdot10^{-3}$ & ...          \vspace{1mm}  \\
DoAr 44             & ... & 100$-$300 & 2.2 &  0.4  & $3.0\cdot10^{-4}$ & $8.8\cdot10^{-5}$ & $1.9\cdot10^{-4}$ & ...          \vspace{1mm}  \\
\underline{RNO 90}  & ... & 100$-$300 & 2.3 &  0.4  & $1.1\cdot10^{-4}$ & $3.1\cdot10^{-5}$ & $7.1\cdot10^{-5}$ & ...          \vspace{1mm}  \\
Wa Oph 6            & 275 & 175$-$375 & 2.4 &  0.7  & $4.9\cdot10^{-4}$ & $8.0\cdot10^{-5}$ & $9.8\cdot10^{-5}$ & ...          \vspace{1mm}    \\
AS 209              & 200 & 100$-$300 & 2.4 &  0.7  & $7.9\cdot10^{-4}$ & $1.2\cdot10^{-4}$ & $1.4\cdot10^{-4}$ & ...          \vspace{1mm}   \\

\hline
\\
\end{tabular}

\label{tab:fits_results}
\begin{flushleft}
1) Underlined objects are those which have not been mapped to date through high-angular resolution imaging. The objects which have been mapped but do not have an estimate for the outer disk radius as reported in Column (2) have been spatially resolve by Andrews et al.~(\cite{And09}). Contrary to Andrews \& Williams~(\cite{And07a}), they modeled the disk surface brightness by using a self-similar profile instead of a truncated power-law. For this reason no estimate for $R_{\rm{out}}$ could be extracted for these sources (see footnote in Sect.~4.1). 
2) Best-fit estimate of the disk outer radius by fitting the observed visibilities at sub-millimeter wavelengths using a truncated power-law for the surface density profile. AW07: Andrews \& Williams~(\cite{And07a}).  
3) Interval of the disk outer radius adopted for our analysis. 
4) Best-fit estimate of the spectral index of the SED $\alpha$ between 1 and 3~mm derived by considering for the outer disk radius the central value of the interval reported in Column (3), and for the power-law index of the surface density profile $p=1$.  
5) Best-fit estimate of the spectral index of the dust opacity $\beta$ between 1 and 3~mm.
6) Product between the dust mass and the dust opacity at 1~mm from the best-fit two-layer disk model.
7) Dust mass obtained with a power-law index for the grain size distribution $q=2.5$. 
8) Like Column (7) but with $q=3$. The sources without an estimate of the dust mass have a $\beta$-value (reported in Column (6)) which cannot be reproduced with $q=3$. 9) Like Column (8) but with $q=3.5$.    
\end{flushleft}
\end{table*}

Even if the estimate for $M_{\rm{dust}}$ depends on the value of $q$, the range spanned by our sample for $q=2.5$ and $q=3$ turns out to be very similar\footnote{Note however that for the four disks with the lowest value of $\beta$ ($\approx 0$) $M_{\rm{dust}}^{q=3}$ cannot be obtained. The reason for this is that these very low $\beta$-values cannot be explained with $q=3$ (see discussion in Sect.~5.1). For the same reason only two disks have an estimate for the dust mass with $q=3.5$.}, 
namely $\sim 2 \cdot 10^{-5} - 3 \cdot 10^{-3}$~$M_{\odot}$, 
corresponding to roughly $6-1000$~$M_{\oplus}$. By defining a planetesimal as a rocky body with a radius of 10~km and a density similar to the one adopted for our dust grain model (i.e. $\rho \sim 1$ g$/$cm$^{3}$), the maximum number of planetesimals which can be potentially formed out of this reservoir of small grains (see discussion below) is $\sim 10^{10} - 10^{12}$.

These numbers have to be taken with great caution mainly because of the large uncertainty for the inferred dust mass of a factor as large as 10 (see e.g. the discussion in Natta et al.~\cite{Nat04}).

Here it is important to remember that observations at (sub-)mm wavelengths are completely insensitive to pebbles/stones much larger than $\sim 1-10$~cm, since the dust opacity decreases as the $a_{\rm{max}}$ of the dust population increases at sizes larger than the wavelengths of the observations. For this reason the dust masses presented here have to be interpreted as lower limits for the real total mass in solids, since in principle large pebbles/stones or even larger rocky bodies like planetesimals may be already present in the disk.

Finally we have investigated relations between dust properties in disks (i.e. $\beta$, $M_{\rm{dust}}$) and the stellar ones (listed in Table~2) but we did not find any significant correlation, similarly to the case of Taurus (R10). No significant correlation was found neither between grain growth and dust mass.

\section{Summary}
\label{sec:summary}

We have presented new observations at $\sim$ 3~mm obtained with the ATCA array and the new CABB digital filter bank for 27 protoplanetary disks in the $\rho-$Oph star forming region. Among these we selected the 17 isolated class II YSOs with well characterized stellar properties (see selection criteria in Sect.~3.1). Our sample comprises all the $\rho-$Oph isolated class II YSOs with an observed flux at 1.3~mm larger than $\sim 75$~mJy, and $\sim 50\%$ of the isolated PMS stars with mass larger than $\sim 0.5$~$M_{\odot}$. We have analyzed the (sub-)millimeter SED of our disk sample and here is the summary of our main findings:  

\begin{enumerate}
 \item The spectral index $\beta$ of the millimeter dust opacity turns out to be lower than the typical value found for the ISM for all the 17 disks detected at 3~mm. The mean value is $<\beta> \approx 0.5$. For the ten disks which have been observed and spatially resolved through past high-angular resolution continuum imaging at sub-mm wavelengths this represents evidence for the presence of dust grains as large as at least $\sim 1$~mm in the disk outer regions. For the seven disks which have not been mapped yet the observations could in principle be consistent also with very compact ($R_{\rm{out}} < 20-30$ AU) disks, significantly different from those mapped so far.
 \item From a comparison between the results found for our sample in $\rho-$Oph and an homogeneously selected sample of 21 isolated class II disks in Taurus-Auriga (R10), there is no statistical evidence of any difference between the distribution of $\beta$-values found in the two star forming regions. This may suggest that environmental effects do not play an important role in the first phases of planet formation.
 \item There is no evidence for any evolution of the dust spectral index: dust grains appear to be present in the outer regions of protoplanetary disks throughout all the class II evolutionary stage of YSOs, confirming what previously found in Taurus. In order to explain the retention of large dust grains in the outer disk some mechanisms which may halt the inward drift of solid particles, e.g. local pressure maxima due to turbulent vortices or spiral density waves, have to be invoked. Since evidence for grain growth to millimeter sizes appears to be present in some class 0 YSOs the formation of the $\sim$ mm-sized grains seen in class II disks may already occur in the densest inner regions of molecular cloud cores (Ormel et al.~\cite{Orm09}).
 \item The mm slope of the SED does not correlate with indicators of dust processing from the silicate feature observed at about 10~$\mu$m, which are sensitive to grain growth to micron sizes in the surface layers of the inner disk; further observations at (sub-)mm wavelengths are needed to extend the investigation to a larger sample.
 \item The spanned range in dust mass contained in grains with sizes $\simless 1$~cm as derived with the dust model described in Sect.~4.2 is about $2\cdot 10^{-5} - 3\cdot 10^{-3}$~$M_{\odot}$ or roughly $6-1000$~$M_{\oplus}$. This reservoir of small grains is capable of forming about $10^{10}-10^{12}$~10 km-sized planetesimals with a mean density of $\sim 1$ g/cm$^3$.
\end{enumerate}

\begin{acknowledgements}
We wish to thank the support astronomers in Narrabri, in particular James Urquhart and Maxim Voronkov, for their help during the ATCA observations. L.R. aknowledges the PhD fellowship of the International Max-Planck-Research School. L.T. and A.N. were partly supported by the grant ASI-COFIS I/016/07/0.
\end{acknowledgements}



\end{document}